\documentclass{article}

\usepackage{amssymb}
\usepackage{graphicx}

\begin{document}

\title{
\textbf{Critical attractors and }$q$\textbf{-statistics\bigskip }}
\author{A. Robledo} 
\date{.}
\maketitle

\begin{center}
Instituto de F\'{\i}sica,Universidad Nacional Aut\'onoma de M\'exico. \\
Apdo Postal 20-364. 01000 M\'exico D.F. Mexico \\
robledo@fisica.unam.mx

\medskip
\textbf{Europhysics News, Vol 36, pages 214-218 (2005)}  
\end{center}

\bigskip

The singular dynamics at critical attractors of even the simplest one
dimensional nonlinear iterated maps is of current interest to statistical
physicists because it provides insights into the limits of validity of the
Boltzmann-Gibbs (BG) statistical mechanics. This dynamics also helps inspect
the form of the possible generalizations of the canonical formalism when its
crucial supports, phase space mixing and ergodicity, break down.

The fame of the critical attractors present at the onset of chaos in the
logistic and circle maps stems from their universal properties, comparable
to those of critical phenomena in systems with many degrees of freedom. At
these attractors the indicators of chaos withdraw, such as the fast rate of
separation of initially close by trajectories. As it is generally
understood, the standard exponential divergence of trajectories in chaotic
attractors suggests a mechanism to justify the property of irreversibility
in the BG statistical mechanics \cite{note1}. In contrast, the onset of
chaos imprints memory preserving properties to its trajectories.

The dynamical nature of trajectories is appraised on a regular basis through
the sensitivity to initial conditions $\xi _{t}$, defined as 
\begin{equation}
\xi _{t}(x_{0})\equiv \lim_{\Delta x_{0}\rightarrow 0}(\Delta x_{t}/\Delta
x_{0}),\;1\ll t,  \label{sensitivity}
\end{equation}%
where $\Delta x_{0}$ is the initial separation of two trajectories and $%
\Delta x_{t}$ that at time $t$. For a one-dimensional map it has the form $%
\xi _{t}(x_{0})=\exp (\lambda _{1}t)$, with $\lambda _{1}>0$ for chaotic
attractors and $\lambda _{1}<0$ for periodic ones. The number $\lambda _{1}$
is called the Lyapunov coefficient. At critical attractors $\lambda _{1}=0$
and $\xi _{t}$ does not settle onto a single-valued function but exhibits
instead fluctuations that grow indefinitely. For initial positions on the
attractor $\xi _{t}$ develops a universal self-similar temporal structure
and its envelope grows with $t$ as a power law.

It has been recently corroborated \cite{robledo1}-\cite{robledo4} that the
dynamics at the critical attractors associated to the three familiar routes
to chaos, intermittency, period doubling and quasiperiodicity \cite{hilborn1}%
, obey the features of the $q$-statistics, the generalization of BG
statistics based on the $q$-entropy $S_{q}$ \cite{tsallis1}. The focal point
of the $q$-statistical description for the dynamics at such attractors is a $%
\xi _{t}$ associated to one or several expressions of the form

\begin{equation}
\xi _{t}(x_{0})=\exp _{q}[\lambda _{q}(x_{0})\ t],  \label{sensitivity1}
\end{equation}%
where $q$ is the entropic index and $\lambda _{q}$ is the $q$-generalized
Lyapunov coefficient. Also the identity $K_{1}=$ $\lambda _{1}$ (where the
rate of entropy production $K_{1}$ is given by $K_{1}t=S_{BG}(t)-S_{BG}(0)$
with $S_{BG}$ the Boltzmann-Gibbs entropy) generalizes to 
\begin{equation}
K_{q}=\lambda _{q},  \label{q-pesin1}
\end{equation}%
where the rate of $q$-entropy production $K_{q}$ is defined via 
$K_{q}t=S_{q}(t)-S_{q}(0)$ \cite{tsallis1}, \cite{robledo2}, \cite{robledo3}.

\bigskip 
\bigskip

\textbf{Tsallis }$q$\textbf{\ index \& Mori's }$q$\textbf{-phase
transitions\bigskip }

The central issue of research in $q$-statistics is perhaps to confirm the
occurrence of special values for the entropic index $q$ for any given system
and to establish their origin. In the case of critical attractors the
allowed values for $q$ are obtained from the universality class parameters
to which the attractor belongs. For the simpler cases, the pitchfork and the
tangent bifurcations, there is a single well-defined value for the index $q$ 
\cite{robledo1}. The pitchfork bifurcations form the sequence of period
doublings that culminate in the chaos threshold whereas at the tangent
bifurcation chaos develops via intermittency. For critical multifractal
attractors, as in the period doubling accumulation point and in the
quasiperiodic onset of chaos, the situation is more complicated and there
appear to be a multiplicity of indexes $q$ but with precise values given by
the attractor scaling functions. They come out in pairs and are related to
the occurrence of dynamical '$q$-phase' transitions \cite{robledo3}, and
these are identified as the source of the special values for the entropic
index $q$. The $q$-phase transitions connect qualitatively different regions
of the multifractal attractor.

The main quantities in the thermodynamic formalism of $q$-phase transitions
(developed by Mori and colleagues in the late 80's \cite{mori1}) are the
spectral functions $\phi (\mathsf{q})$ and $\psi (\lambda )$, related to
each other via Legendre transformation, the function of generalized Lyapunov
coefficients $\lambda (\mathsf{q})$, given by $\lambda (\mathsf{q})\equiv
d\phi (\mathsf{q})/d\mathsf{q}$, and the variance $v(\mathsf{q})$ $\equiv
d\lambda (\mathsf{q})/d\mathsf{q}$. The functions $\phi (\mathsf{q})$ and $%
\psi (\lambda )$ are the dynamic counterparts of the multifractal dimensions 
$D(\mathsf{q})$ and the spectrum $f(\alpha )$ that characterize the
geometric structure of the attractor \cite{hilborn1}. As with ordinary
thermal 1st order phase transitions, a $q$-phase\ transition is indicated by
a section of linear slope $m_{c}=1-q$ in the 'free energy' $\psi (\lambda )$%
, a discontinuity at $\mathsf{q}=q$ in the 'order parameter' $\lambda (%
\mathsf{q})$, and a divergence at $q$ in the 'susceptibility' $v(\mathsf{q})$%
. Actually an infinite family of such transitions takes place but of rapidly
decreasing strengths \cite{robledo3}, \cite{robledo4}.\bigskip

$\bullet $ \textbf{Tangent \& pitchfork bifurcations\bigskip }

The tangent bifurcations of unimodal (one hump) maps of general nonlinearity 
$z>1$ display weak insensitivity to initial conditions, i.e. power-law
convergence of orbits when at the left-hand side ($x<x_{c}$) of the point of
tangency $x_{c}$. However at the right-hand side ($x>x_{c}$) of the
bifurcation there is a `super-strong' sensitivity to initial conditions,
i.e. a sensitivity that grows faster than exponential \cite{robledo1}. The
two different behaviors can be couched as a $q$-phase transition with
indexes $q$ and $2-q$ for the two sides of the tangency point. The pitchfork
bifurcations display weak insensitivity to initial conditions.

For the tangent bifurcations one has always $q=3/2$, while for the pitchfork
bifurcation one has $q=5/3$. Notably, these results for the index $q$ are
valid for all $z>1$ and therefore define the existence of only two
universality classes for unimodal maps \cite{robledo1}. The treatment of the
tangent bifurcation differs from other studies of intermittency transitions
in that there is no feed back mechanism of iterates from an adjacent chaotic
region. Therefore, impeded or incomplete mixing in phase space (a small
interval neighborhood around $x=x_{c}$) arises from the special 'tangency'
shape of the map at the transitions that produces monotonic trajectories.
This has the effect of confining or expelling trajectories causing anomalous
phase-space sampling, in contrast to the thorough coverage in generic states
with $\lambda _{1}>0$.\bigskip

$\bullet $ \textbf{Period-doubling accumulation point\bigskip }

For a unimodal map of nonlinearity $z>1$ (e.g. the logistic map has $z=2$)
with extremum at $x=0$ and control parameter $\mu $ the onset of chaos is
obtained at the accumulation point $\mu _{\infty }$ of the $\mu $ values for
the pitchfork bifurcations $\mu _{n}$, $n=1,2,...$, \cite{hilborn1}. This is
often called the Feigenbaum attractor which reappears in multiples together
with the precursor cascade of period-doubling bifurcations in the infinite
number of windows of periodic trajectories that interpose the chaotic
attractors beyond $\mu _{\infty }$. The number of cascades within each
window is equal to the period of the orbit that emerges at the tangent
bifurcation at its opening. See Fig. 1. The dynamics at the Feigenbaum
attractor has been analyzed recently \cite{robledo2}, \cite{robledo3}. By
taking as initial condition $x_{0}=0$ at $\mu _{\infty }$ it is found that
the resulting orbit of period $2^{\infty }$ consists of trajectories made of
intertwined power laws that asymptotically reproduce the entire
period-doubling cascade that occurs for $\mu <\mu _{\infty }$. See Fig. 2a.
It was established that $\xi _{t}$ has precisely the form of a set of
interlaced $q$-exponentials, of which the $q$-indexes and the sets of
associated $q$-Lyapunov coefficients $\lambda _{q}$ were determined. As
mentioned, the appearance of a specific value for the $q$ index (and
actually also that for its conjugate value $Q=2-q$) turns out to be due to
the occurrence of $q$-phase transitions between 'local attractor structures'
at $\mu _{\infty }$. The values of the $q$-indexes are obtained from the
discontinuities of the universal trajectory scaling function $\sigma $. This
function characterizes the multifractal by measuring how adjacent positions
of orbits of period $2^{n}$ approach each other as $n\rightarrow \infty $ 
\cite{schuster1}. The main discontinuity in $\sigma $ is related to the most
crowded and most sparse regions of the attractor and in this case $q=1-\ln
2/(z-1)\ln \alpha _{F}$, where $\alpha _{F}$ is the universal scaling
constant associated to these two regions. Furthermore, it has also been
shown \cite{robledo2}, \cite{robledo3} that the dynamical and entropic
properties at $\mu _{\infty }$ are naturally linked through the $q$%
-exponential and $q$-logarithmic expressions, respectively, for the
sensitivity $\xi _{t}$ and for the entropy $S_{q}$ in the rate of entropy
production $K_{q}$.\textbf{\bigskip }

$\bullet $ \textbf{Quasiperiodicity \& golden mean route to chaos\bigskip }

A recent study \cite{robledo4} of the dynamics at the quasiperiodic onset of
chaos in maps with zero slope inflection points of cubic nonlinearity (e.g.
the critical circle map) shows strong parallelisms with the dynamics at the
Feigenbaum attractor described above. Progress on detailed knowledge about
the structure of the orbits within the golden-mean quasiperiodic attractor,
see Fig. 2b, helped determine the sensitivity to initial conditions for sets
of starting positions within this attractor \cite{robledo4}. It was found
that $\xi _{t}$ is made up of a hierarchy of families of infinitely many
interconnected $q$-exponentials. Here again, each pair of regions in the
multifractal attractor, that contain the starting and finishing positions of
a set of trajectories, leads to a family of $q$-exponentials with a fixed
value of the index $q$ and an associated spectrum of $q$-Lyapunov
coefficients $\lambda _{q}$.

As in the period doubling route to chaos, the quasiperiodic dynamics
consists of an infinite family of $q$-phase transitions, each associated to
trajectories that have common starting and finishing positions located at
specific regions of the multifractal. The specific values of the variable $%
\mathsf{q}$ (in the thermodynamic formalism) at which the $q$-phase
transitions take place are the same values for the entropic index $q$ in $%
\xi _{t}$. The transitions come in pairs at $q$ and $2-q$ as they are tied
down to the expressions for $\lambda _{q}$ in $\xi _{t}$. Again, the
dominant dynamical transition is associated to the most crowded and sparse
regions of the multifractal, and the value of its $q$-index is \cite%
{robledo4} $q=1-\ln w_{gm}/2\ln \alpha _{gm}$ where $w_{gm}=(\sqrt{5}-1)/2$
is the golden mean and $\alpha _{gm}$ is the universal constant that plays
the same scaling role as $\alpha _{F}$.\bigskip

$\bullet $ \textbf{Structure in dynamics\bigskip\ }

The dynamical organization within critical multifractal attractors is
difficult to resolve from the consideration of a straightforward time
evolution, i.e. the record of positions at every time $t$ for a trajectory
started at an arbitrary position $x_{0}$ within the attractor. In this case
what is observed are strongly fluctuating quantities that persist in time
with a scrambled pattern structure that exhibits memory retention.
Unsystematic averages over $x_{0}$ would rub out the details of the
multiscale dynamical properties we uncovered. On the other hand, if specific
initial positions with known location within the multifractal are chosen,
and subsequent positions are observed only at pre-selected times, when the
trajectories visit another selected region, a distinct $q$-exponential
expression for $\xi _{t}$ is obtained.\bigskip

\textbf{Manifestations of }$q$\textbf{-statistics in condensed matter
problems\bigskip }

There are connections between the properties of critical attractors
referred-to here and those of systems with many degrees of freedom at
extremal or transitional states. Three specific examples have been recently
developed. See Table. In one case the dynamics at the chaos threshold via
intermittency has been shown to be related to that of intermittent clusters
at thermal critical states \cite{robcrit1}. In the second case the dynamics
at the noise-perturbed period-doubling onset of chaos has been found to show
parallelisms with the glassy dynamics observed in supercooled molecular
liquids \cite{robglass1}. In the third case the known connection between the
quasiperiodic route to chaos and the localization transition for transport
in incommensurate systems is analyzed from the perspective of the $q$%
-statistics \cite{roblocn1}.\bigskip

$\bullet $ \textbf{Critical clusters \& intermittency\bigskip }

The dynamics of fluctuations of an equilibrium critical state in standard
models of a magnetic or fluid system is seen to be related to the dynamics
at a critical attractor for intermittency. The connection can be examined
when instead of the entire critical system a single unstable cluster of
excess magnetization or density of size $R$ is considered. The analysis,
initially developed by Contoyiannis and colleagues \cite{athens1}, has been
reconsidered recently in connection to $q$-statistics \cite{robcrit1}.

An important element in the analysis is the determination of the order
parameter $\phi (${\boldmath$r$}$)$ of a large cluster by withholding only
its most probable configurations from a coarse-grained partition function $Z$%
. The conditions under which these configurations dominate are evaluated as
these determine an instability of the cluster. The instability can be
expressed as an inequality between two lengths in space. This is $r_{0}\gg R$%
, where $r_{0}$ is the location of a divergence in the expression for $\phi
( ${\boldmath$r$}$)$. In a coarse-grained time scale the cluster is expected
to evolve by increasing its average amplitude $\overline{\phi }$ and/or size 
$R$ because the subsystem studied represents an environment with unevenness
in the states of the microscopic degrees of freedom (e.g. more spins up than
down). Increments in $\overline{\phi }$ for fixed $R$ takes the position $%
r_{0}$ for the singularity closer to $R$, the dominance of this
configuration in $Z$ decreases accordingly and rapidly so. When the
divergence is reached at $r_{0}=$ $R$ the profile $\phi (r)$ no longer
describes the spatial region where the subsystem is located. But a
subsequent fluctuation would again be represented by a cluster $\phi (r)$ of
the same type. From this renewal process we obtain a picture of
intermittency \cite{robcrit1}.

Amongst the static and dynamical properties for such single critical cluster
of order parameter $\phi (${\boldmath$r$}$)$ we mention \cite{robcrit1}: 1)
The faster than exponential growth with cluster size $R$ of the
space-integrated $\phi $ suggests nonextensivity of the BG entropy but
extensivity of a $q$-entropy expression. 2) The finding that the time
evolution of $\phi $ is described by the dynamics of the critical attractor
for intermittency which implies an atypical sensitivity to initial
conditions compatible with $q$-statistics. 3) Both, the approach to
criticality and the infinite-size cluster limit at criticality manifest
through a crossover from canonical to $q$-statistics.\bigskip

$\bullet $ \textbf{Glassy dynamics \& noise-perturbed Feigenbaum
attractor\bigskip }

The erratic motion of a Brownian particle is usually described by the
Langevin theory. As it is well known, this method finds a way round the
detailed consideration of many degrees of freedom by representing the effect
of collisions with molecules in the fluid in which the particle moves by a
noise source. The approach to thermal equilibrium is produced by random
forces, and these are sufficient to determine dynamical correlations,
diffusion, and a basic form for the fluctuation-dissipation theorem. In the
same spirit, attractors of nonlinear low-dimensional maps under the effect
of external noise can be used to model states in systems with many degrees
of freedom. In a one-dimensional map with only one control parameter $\mu $
the consideration of external noise of amplitude $\sigma $ could be thought
to represent the effect of many other systems coupled to it, like in the
so-called coupled map lattices \cite{schuster1}. The general map formula can
be seen to represent a discrete form for a Langevin equation with nonlinear
`friction force' term \cite{robglass1}.

The dynamics of noise-perturbed logistic maps at the chaos threshold has
been shown to exhibit parallels with the most prominent features of glassy
dynamics in, for example, supercooled liquids. In this analogy the noise
amplitude $\sigma $ plays a role similar to the temperature difference from
a glass transition temperature. Specifically our results are \cite{robglass1}%
: 1) Two-step relaxation occurring in trajectories and in their two-time
correlations when $\sigma \rightarrow 0$. 2) A map equivalent to a
relationship between the relaxation time and the configurational entropy. 3)
Both, trajectories and their two-time correlations obey an `aging' scaling
property typical of glassy dynamics when $\sigma =0$. 4) A progression from
normal diffusiveness to subdiffusive behavior and finally to a stop in the
growth of the mean square displacement as demonstrated by the use of a
repeated-cell map. See Fig. 3. The existence of this analogy is perhaps not
accidental since the limit of vanishing noise amplitude $\sigma \rightarrow
0 $ involves loss of ergodicity.\bigskip

$\bullet $ \textbf{Localization \& quasiperiodic onset of chaos\bigskip }

One interesting problem in condensed matter physics that exhibits
connections with the quasiperiodic route to chaos is the localization
transition for transport in incommensurate systems, where the discrete Schr%
\"{o}dinger equation with a quasiperiodic potential translates into a
nonlinear map known as the Harper map \cite{satija1}. In this equivalence
the divergence of the localization length corresponds to the vanishing of
the ordinary Lyapunov coefficient. It is interesting to note that the basic
features of $q$-statistics in the dynamics at critical attractors mentioned
here turn up in the context of localization phenomena.\bigskip 

\textbf{Acknowledgements.} Partially supported by CONACyT and DGAPA-UNAM,
Mexican agencies.

\begin{table}
\begin{center}
\begin{tabular}[c]{|l|l|l|l|}
\hline
{\bf Route to chaos}&
{\bf Intermittency}&
{\bf Period doubling}&
{\bf Quasiperiodicity}\\
\hline
{\bf Common}&
\multicolumn{3}{l|}{Vanishing ordinary Lyapunov coefficient,}\\
{\bf properties}&
\multicolumn{3}{l|}{dynamical phase transitions (Mori's $q$-phases),}\\
&
\multicolumn{3}{l|}{power-law dynamics, $q$-sensitivity, $q$-Pesin
identity}\\
\hline
{\bf Distinctive}&
(Also) faster&
Foam-like&
Dense\\
{\bf properties}&
than exponential&
phase space&
phase space\\
&
dynamics&
&\\
\hline
{\bf Applications in}&
Critical&
Glass&
Localization\\
{\bf condensed}&
clusters&
formation&
\\
{\bf matter physics}&
&
&
\\
\hline
{\bf Applications in}&
Information \&&
Protein folding,&
Mode locking,\\
{\bf other}&
other flows&
vegetation&
cardiac cells,\\
{\bf disciplines}&
in networks,\ldots&
patterns,\ldots&
Internet TCP\\
\hline
\end{tabular}
\end{center}
\caption{Summary. The three routes to chaos, properties and applications.}
\end{table}

\newpage

\begin{figure}
\includegraphics[width=1.0\columnwidth,angle=-90,scale=0.6]{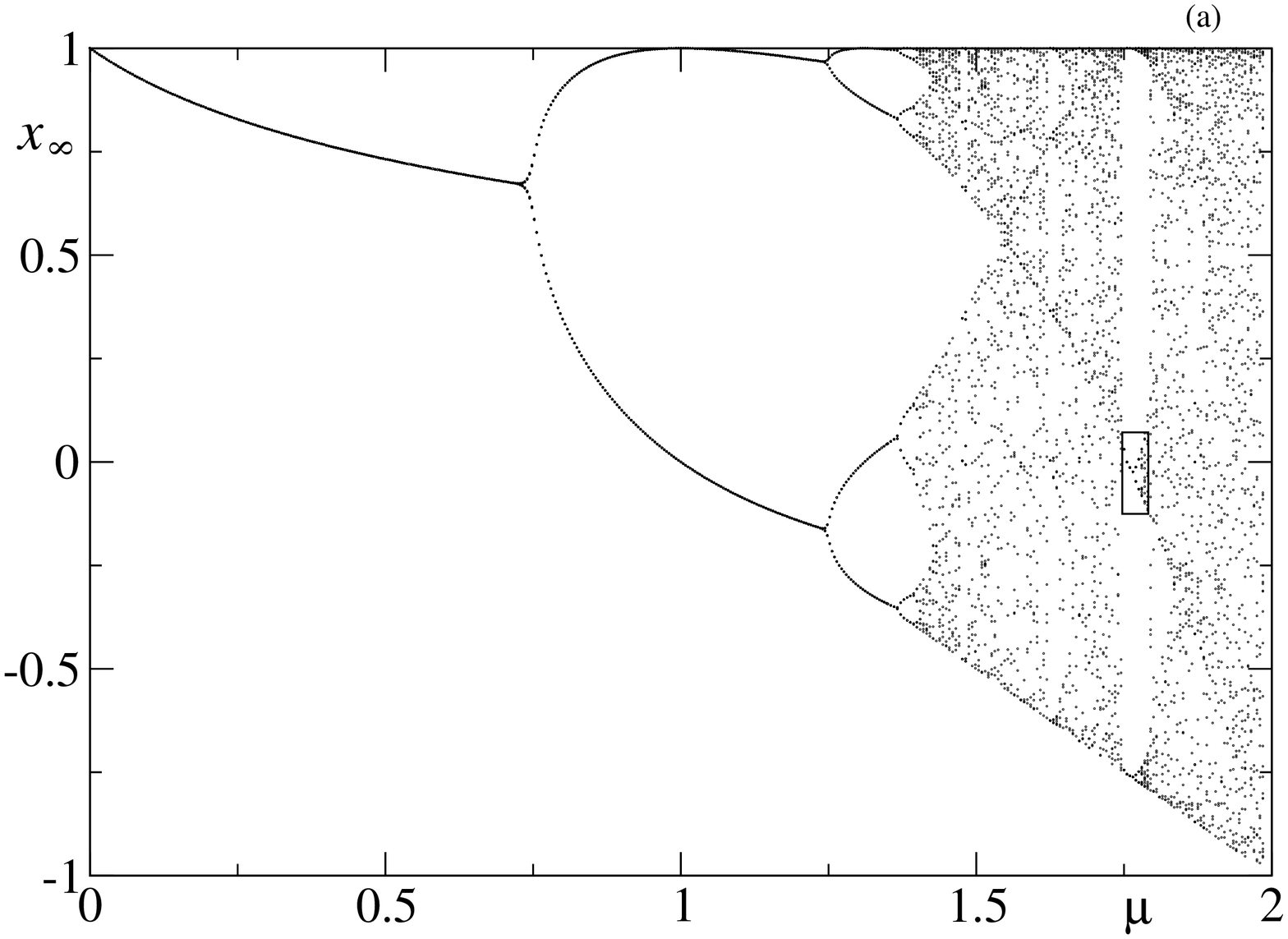}
\end{figure}

\begin{figure}
\includegraphics[width=1.0\columnwidth,angle=-90,scale=0.6]{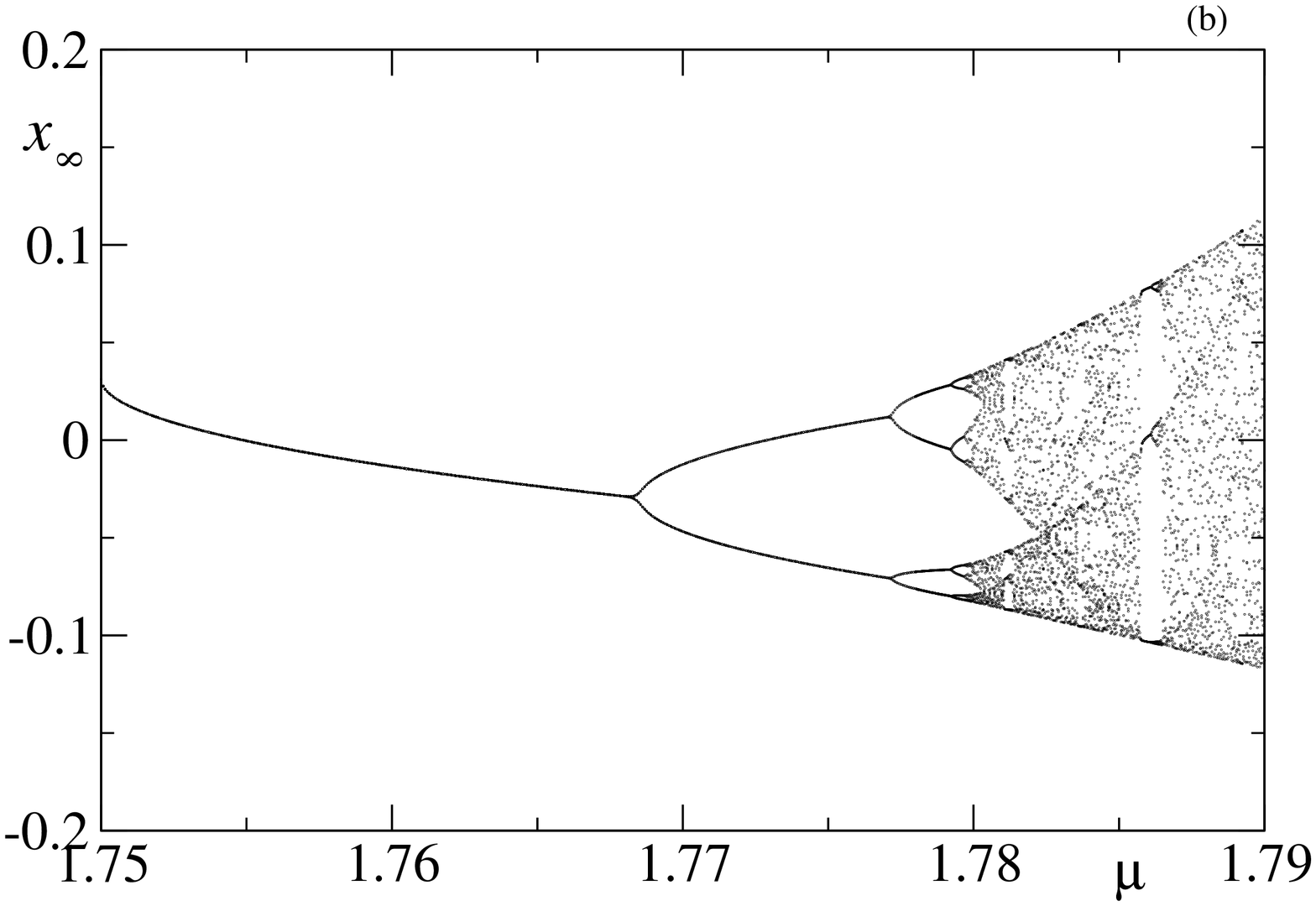}
\caption{(a) The classic logistic map attractor as a function of control
parameter $\mu $. (b) Enlargement of the box in (a).}
\label{figure1}
\end{figure}

\begin{figure}
\includegraphics[width=1.0\columnwidth,angle=0,scale=0.7]{fig_2a.eps}
\includegraphics[width=1.0\columnwidth,angle=-90,scale=0.7]{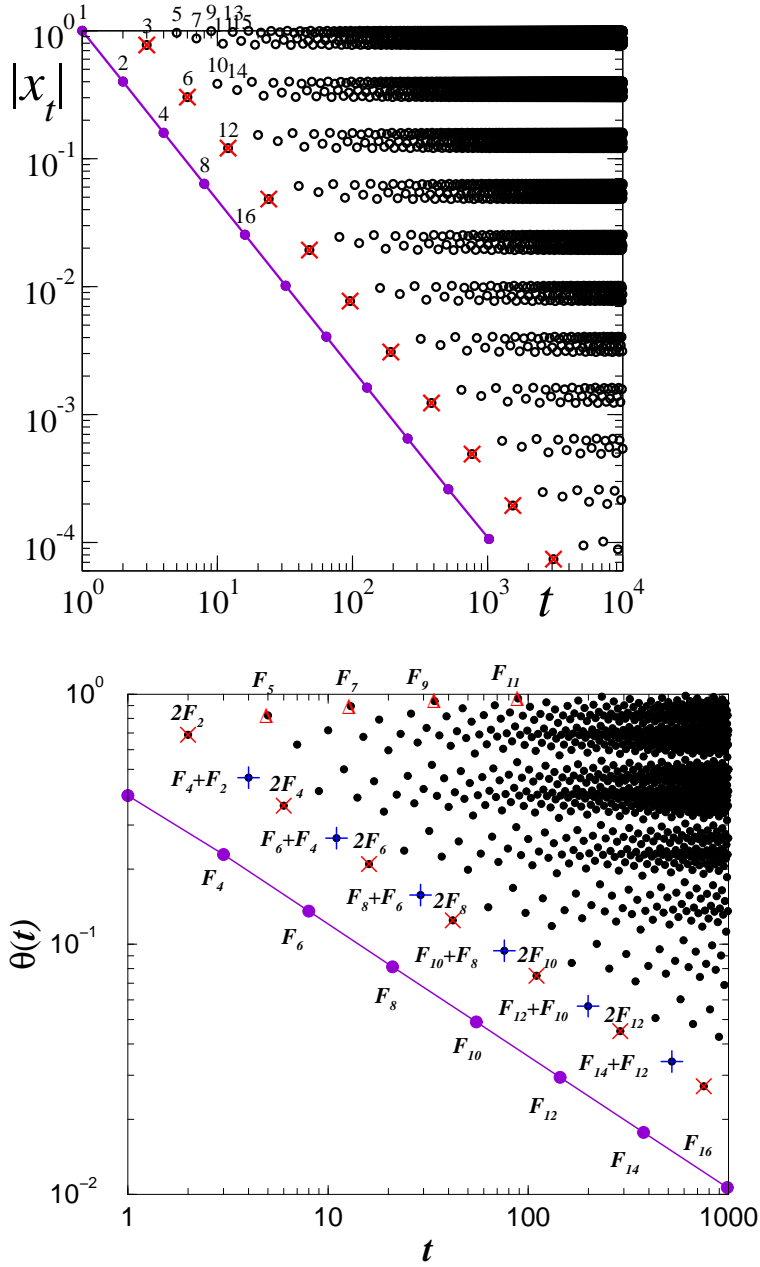}
\caption{(a) Absolute values of positions in logarithmic scales of the first
10000 iterations for the trajectory of the logistic map at the onset of
chaos $\mu _{\infty }$ with initial condition $x_{0}=0$. The numbers
correspond to iteration times. The power-law decay of time subsequences
mentioned in the text can be clearly appreciated. (b) Positions $\theta (t)$
vs $t$ in logarithmic scales for the orbit with initial condition $\theta
(0)=0$ of the critical circle map for the golden-mean winding number. The
labels indicate iteration time $t$ where $F_{n}$ is the Fibonacci number of
order $n$.}
\label{figure2}
\end{figure}

\begin{figure}
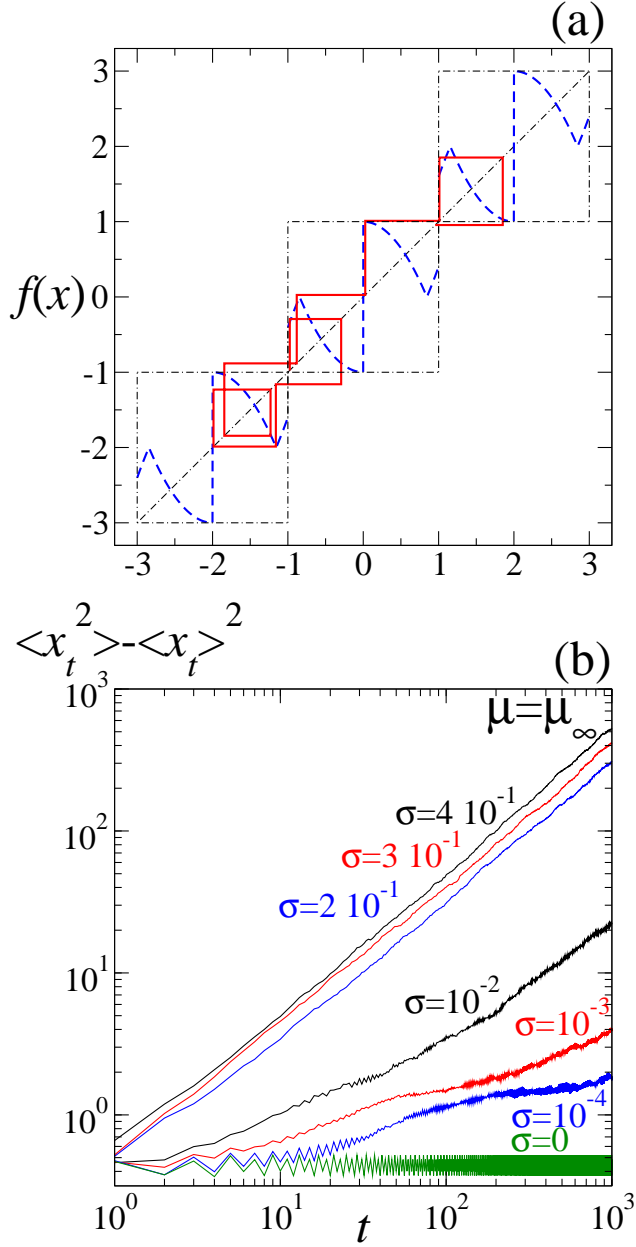

\includegraphics[width=1.0\columnwidth,angle=0,scale=0.7]{fig_3a.eps}
\includegraphics[width=1.0\columnwidth,angle=0,scale=0.7]{fig_3b.eps}
\caption{Glassy diffusion in the noise-perturbed logistic map. (a)
Repeated-cell map (blue) and trajectory (red). (b) Time evolution of the
mean square displacement $\left\langle x_{t}^{2}\right\rangle -\left\langle
x_{t}\right\rangle ^{2}$ for an ensemble of $1000$ trajectories with initial
conditions randomly distributed inside $[-1,1]$. Curves are labeled by the
value of the noise amplitude.}
\label{figure3}
\end{figure}


\begin{thebibliography}{99}
\bibitem{note1} There are some parallels between the evolution towards an
attractor in dissipative maps and the irreversible approach to equilibrium
in thermal systems. It should be recalled that for the chaotic dynamics of a
conservative system (such as the hard sphere gas) there is no dissipation
nor strange attractor. It is of course this dynamics that is relevant to
Boltzmann's assumption of molecular chaos.
\end{thebibliography}
\end{document}